\documentclass[showpacs,showkeys,preprintnumbers]{revtex4}

\usepackage{graphicx}
\usepackage{latexsym}
\usepackage{amssymb}

\def\beq{\begin{equation}}
\def\eeq{\end{equation}}
\def\rmd{{\rm d}} 
\def\rmD{{\rm D}}

\def\pmb#1{\setbox0=\hbox{$#1$}%
  \kern-.025em\copy0\kern-\wd0
  \kern.05em\copy0\kern-\wd0
  \kern-.025em\raise.0433em\box0}
  
\def\bfphi{\pmb{\phi}}

\begin{document}

\title{Scattering by an electromagnetic radiation field}

\author{D. Bini}
\affiliation{
Istituto per le Applicazioni del Calcolo ``M. Picone,'' CNR, I-00185 Rome, Italy\\
ICRA, University of Rome ``La Sapienza,'' I-00185 Rome, Italy\\
INFN, Sezione di Firenze, I--00185 Sesto Fiorentino (FI), Italy}

\author{A. Geralico}
  \affiliation{Physics Department and
ICRA, University of Rome ``La Sapienza,'' I-00185 Rome, Italy}

\begin{abstract}
Motion of test particles in the gravitational field associated with an electromagnetic plane wave is investigated.
The interaction with the radiation field is modeled by a force term {\it \`a la} Poynting-Robertson entering the equations of motion given by the 4-momentum density of radiation observed in the particle's rest frame with a multiplicative constant factor expressing the strength of the interaction itself.
Explicit analytical solutions are obtained.
Scattering of fields by the electromagnetic wave, i.e., scalar (spin 0), massless spin $\frac12$ and electromagnetic (spin 1) fields, is studied too.
\end{abstract}



\maketitle

\section{Introduction}

The problem of scattering by a radiation field on test particle motion was first investigated long ago by Poynting \cite{Poynting-03} in the context of Newtonian gravity, and then generalized to the case of a weak gravitational field by Robertson \cite{Robertson-37}.
The particles interact with the radiation field of an emitting source by adsorbing and re-emitting radiation, causing a drag force responsible for deviation from geodesic motion, known as Poynting-Robertson effect. 
Recently the generalization to the framework of general relativity has been developed in Refs. \cite{BiniJS-09,BiniGJSS-11}, where this effect on test particles orbiting in the equatorial plane of a Schwarzschild or Kerr spacetime has been considered.
The radiation field is taken there as a test field superposed to the gravitational background.   
In Ref. \cite{vaidyaPR} a self-consistent radiation flux was instead used to investigate such a kind of interaction in the exact, Vaidya spherically symmetric spacetime whose source is a null dust \cite{Vaidya-43,Vaidya-51}. 
Particles undergo in this case the action of a Thomson-type interaction with the radiation in terms of which the energy-momentum tensor is interpreted. The way adopted to model this interaction simply consists in taking the 4-momentum density of radiation observed in the particle's rest frame and fix, by an effective proportionality constant, what part of it is being transferred to the particle.

In the present paper we follow the same approach to study the interaction of test particles with an electromagnetic radiation field.
Suppose that before the passage of the wave the spacetime is flat.
An electromagnetic wave propagating over a spacetime region makes it not empty and not flat. 
Therefore, the spacetime curvature associated with an electromagnetic pulse, namely the associated gravitational field, induces observable effects on test particle motion.
While it seems that the relativistic community is quite familiar with the effects related to the passage of a gravitational wave it is somehow surprising that its electromagnetic counterpart is rather poorly studied.
It is expected that a test particle as well as a test field can be scattered by the wave and hence it will modify its own energy and momentum as a consequence of this interaction.

An exact solution of Einstein's field equations representing the gravitational field associated with an electromagnetic radiation field was discovered long ago by Griffiths \cite{grif}.
He then studied the strong field interaction between exact gravitational and electromagnetric waves \cite{grif2,grif3}.
This work has been further generalized in Refs. \cite{gurtug1,gurtug2}, where colliding wave packets consisting of hybrid mixtures of electromagnetic, gravitational and scalar waves in a strong field regime were considered.
We will study first the geodesics of such a solution, which can be determined fully analytically, so that the first approximation approach to the scattering problem can be given a complete answer. 
We will then investigate the interaction with the radiation field filling the spacetime region by considering accelerated orbits with acceleration proportional to the energy-momentum distribution of the wave. In a sense, in the scattering process the particle absorbs and re-emits radiation, resulting in a force term acting on the particle itself. The equations of motion can be analytically solved also in this case.
We may consider this as a second approximation approach to the scattering problem.
Finally, we will study also fields besides particles as scattered by the electromagnetic wave, namely scalar (spin 0), massless spin $\frac12$ and electromagnetic (spin 1) fields, and discuss the features of the scattering as well as the coupling between quantum numbers and background parameters.

\section{Background spacetime}

An exact solution representing the gravitational field associated with an electromagnetic plane wave propagating along the $z$ axis is given by \cite{grif}
\beq
\label{radfield}
\rmd s^2 = -2\rmd u \rmd v + \cos^2 (bu) (\rmd x^2 + \rmd y^2)\,,
\eeq
with an energy-momentum tensor
\beq
\label{Tmunu}
T=2b^2 k\otimes k \,,\qquad k=\partial_v\,,
\eeq
so that $G_{\mu\nu}=T_{\mu\nu}$.
In the $(u,v,x,y)$ coordinates, the metric (\ref{radfield}) has a horizon at $ bu=\pi/2$; therefore the allowed range for the coordinate $u$ is $[0,u_0]$ with $u_0<\pi/(2b)$.
The metric (\ref{radfield}) is also conformally flat; in fact one can introduce the new variable $u'=(\tan bu)/b$ such that the new form of the metric is
\beq
\label{conf_flat}
\rmd s^2 = \frac{1}{1+(bu')^2}\left(-2\rmd u' \rmd v + \rmd x^2 + \rmd y^2 \right)\,,
\eeq
and no coordinate horizon exist anymore.
Using instead the transformation
\beq
u=\frac{1}{\sqrt{2}}(t-z)\,,\qquad v=\frac{1}{\sqrt{2}}(t+z)\,,\qquad 
\eeq
with $x$ and $y$ unchanged allows to obtain the quasi-Cartesian form of the metric (\ref{radfield})
\beq
\label{quasicart}
\rmd s^2 = -\rmd t^2 + \cos^2 s \,(\rmd x^2 + \rmd y^2)+ \rmd z^2\,.
\eeq
where we have denoted $\omega={b}/{\sqrt{2}}$ and $s=\omega (t-z)$.
Hereafter we will work with the metric in the form (\ref{quasicart}), which reduces to the flat Cartesian spacetime metric in the limit $\omega \to 0$ (or $s\to 0$).

A family of fiducial observers at rest with respect to the coordinates $(x,y,z)$ is characterized by the $4$-velocity vector
\beq
m=\partial_t\,,\qquad m^\flat =-\rmd t\,.
\eeq
An orthonormal spatial triad adapted to the observers $m$ is given by
\beq
\label{ort_frame}
e_{\hat x}=\frac{1}{\cos s}\partial_x\,,\quad 
e_{\hat y}=\frac{1}{\cos s}\partial_y\,,\quad 
e_{\hat z}=\partial_z\,,
\eeq
with dual
\beq
\label{ort_frame_forms}
\omega^{\hat x}= \cos s\, \rmd x\,,\quad 
\omega^{\hat y}= \cos s \, \rmd y\,,\quad 
\omega^{\hat z}=\rmd z\,.
\eeq
The associated  congruence of the observers world lines is geodesic and vorticity-free but has a nonzero expansion
\beq
\theta(m)=-\omega \tan s \,[\omega^{\hat x} \otimes \omega^{\hat x}+\omega^{\hat y} \otimes \omega^{\hat y}]\,,\quad \Theta(m)={\rm Tr}[\theta(m)]=-2\omega \tan s\,.
\eeq
The electric (${\mathcal E}(m)$), magnetic (${\mathcal H}(m)$) and mixed (${\mathcal F}(m)$) parts of the Riemann tensor (see, e.g., Ref. \cite{book} for their standard definitions) have constant frame components  given by
\beq
{\mathcal E}(m)=\omega^2 \,[\omega^{\hat x} \otimes \omega^{\hat x}+\omega^{\hat y} \otimes \omega^{\hat y}]\,, \qquad
{\mathcal H}(m)=-\omega^2 \,\omega^{\hat x} \wedge\omega^{\hat y}\,,
\eeq
with ${\mathcal E}(m)={\mathcal F}(m)$ as for a conformally flat spacetime.

As seen by the observers $m$, a test particle in motion with $4$-velocity $U^\alpha=\rmd x^\alpha/\rmd \tau$ is such that
\beq
\label{Udef}
U=U^\alpha \partial_\alpha =\gamma(U,m) [m +\nu(U,m)^{\hat a} e_{\hat a}]
=\gamma(U,m) [m +||\nu(U,m)||\hat \nu(U,m)]\,,
\eeq
leading to the following relation between coordinate and frame components 
\beq
\label{Ucompts}
U^t=\gamma(U,m)\,,\quad
\frac{U^x}{U^t }=\frac{\nu(U,m)^{\hat x}}{\cos s} \,,\quad
\frac{U^y}{U^t }=\frac{\nu(U,m)^{\hat y}}{\cos s} \,,\quad
\frac{U^z}{U^t }= \nu(U,m)^{\hat z}\,. 
\eeq
For a test particle moving along a generic timelike geodesic $U_{\rm (g)}$ we find
\begin{eqnarray}
U_{\rm (g)}&=&\frac{1}{\sqrt{2}}
\left[ p_u +\frac{1}{2p_u}\left(1+\frac{p_x^2+p_y^2}{ \cos^2 s}\right)\right]\partial_t+\frac{p_x}{\cos^2 s}\partial_x+
\frac{p_y}{\cos^2 s}\partial_y
+\frac{1}{\sqrt{2}}\left[ -p_u +\frac{1}{2p_u}\left(1+\frac{p_x^2+p_y^2}{ \cos^2 s}\right)\right]\partial_z\,,
\end{eqnarray}
so that there exist three (Killing) constants of motion, i.e., $p_x$, $p_y$, $p_u$ (not to be confused with the covariant components of $U_{\rm (g)}$).
Equivalently, using the notation $p_x=p_\perp \cos\alpha$, $p_y=p_\perp \sin\alpha$, so that  $p_\perp^2=p_x^2+p_y^2$, the frame components of the geodesic $4$-velocity are given by
\begin{eqnarray}
\label{geo}
\gamma(U_{\rm (g)},m)&=& \frac{1}{\sqrt{2}}
\left[ p_u +\frac{1}{2p_u}\left(1+\frac{p_\perp^2}{\cos^2 s}\right)\right]\,,\nonumber\\
\gamma(U_{\rm (g)},m)\nu(U_{\rm (g)},m)^{\hat x}&=&\frac{p_\perp \cos\alpha}{\cos s}\,,\nonumber\\
\gamma(U_{\rm (g)},m)\nu(U_{\rm (g)},m)^{\hat y}&=&\frac{p_\perp \sin\alpha}{\cos s}\,,\nonumber\\
\gamma(U_{\rm (g)},m)\nu(U_{\rm (g)},m)^{\hat z}&=&\frac{1}{\sqrt{2}}\left[ -p_u +\frac{1}{2p_u}\left(1+\frac{p_\perp^2}{\cos^2 s} \right)\right]\,,
\end{eqnarray}
with
\beq
||\nu(U_{\rm (g)},m)||^2=\displaystyle\frac{
\left(p_u-\frac{1}{2p_u}  \right)^2
+\frac{p_\perp^2}{\cos^2 s}\left(
1+\frac{1}{2p_u^2}+\frac{p_\perp^2}{4p_u^2 \cos^2 s} \right) }
{\left(p_u+\frac{1}{2p_u}  \right)^2+\frac{p_\perp^2}{\cos^2 s}\left(1+\frac{1}{2p_u^2}+\frac{p_\perp^2}{4p_u^2 \cos^2 s} \right) }\,.
\eeq
In the special case $p_x=0=p_y$ of motion along the $z$ direction, these relations reduce to
\beq
U_{\rm (g)}=\frac{1}{\sqrt{2}}
\left( p_u +\frac{1}{2p_u} \right) \partial_t +
\frac{1}{\sqrt{2}}\left( -p_u +\frac{1}{2p_u} \right)\partial_z=\gamma_0 (\partial_t +\nu_0 \partial_z)\,,
\eeq
where
\beq
\gamma_0=\frac{1+2p_u^2}{2\sqrt{2}p_u}\,,\qquad \nu_0=\frac{1-2p_u^2}{1+2p_u^2}\,,
\eeq
with $\nu_0>0$ if $p_u<1/\sqrt{2}$ and $\nu_0\le 0$ if $p_u\ge 1/\sqrt{2}$.
It is also useful  to consider the special case $p_u=\sqrt{1+p^2}/\sqrt{2}$, $p_x=p$ and $p_y=0$, which in flat spacetime (i.e., before the passage of the background electromagnetic wave: $s=0$ or $\omega=0$) corresponds to a motion confined along the $x$ axis
\begin{eqnarray}
\label{time_geo_spec}
U_{\rm (g)}^{(x)}&=&  
\left[ \sqrt{1+p^2}+\frac{p^2\tan^2 s}{ 2\sqrt{1+p^2}} \right]\partial_t+\frac{p}{\cos^2 s}\partial_x
+\frac{p^2\tan^2 s}{ 2\sqrt{1+p^2}}  \partial_z\,,
\qquad \lim_{s\to 0}U_{\rm (g)}^{(x)}= \sqrt{1+p^2} \partial_t+p \partial_x\,.
\end{eqnarray}

Finally a null geodesic $K_{\rm (g)}$ has 4-momentum 
\begin{eqnarray}
\label{null_geo}
K_{\rm (g)}&=&\frac{1}{\sqrt{2}}
\left(k_u +\frac{1}{2k_u}\frac{k_x^2+k_y^2}{ \cos^2 s}\right)\partial_t+\frac{k_x}{\cos^2 s}\partial_x+
\frac{k_y}{\cos^2 s}\partial_y
+\frac{1}{\sqrt{2}}\left( -k_u +\frac{1}{2k_u}\frac{k_x^2+k_y^2}{ \cos^2 s}\right)\partial_z\,,
\end{eqnarray}
with $k_u$, $k_x$ and $k_y$ Killing constants (not to be confused with the covariant components of $K_{\rm (g)}$; $k_u>0$ in order $K_{\rm (g)}$ to be future-pointing).
The particular choice $k_u=\Omega/\sqrt{2}$, $k_x=\Omega$ and $k_y=0$ gives
\begin{eqnarray}
\label{null_geo_spec}
K_{\rm (g)}^{(x)}&=&\frac{\Omega}{2}\left(1+\frac{1}{\cos^2 s} \right)\partial_t +\frac{\Omega}{\cos^2 s} \partial_x+
\frac{\Omega}{2}\left(-1+\frac{1}{\cos^2 s} \right)\partial_z\,, \qquad \lim_{s\to 0}K_{\rm (g)}^{(x)}= \Omega (\partial_t+\partial_x)
\,,
\end{eqnarray}
which corresponds in the flat spacetime limit to a motion along the $x$ axis.
We see that a component along the $z$ axis arises due to the transfer of momentum by the wave in that direction, as expected.

\section{Scattering of test particles: Poynting-Robertson-like effect}

Let us consider now orbits accelerated by the background radiation field.
We follow here the same approach adopted in Refs. \cite{BiniJS-09,BiniGJSS-11}, where the effect of the interaction with a test radiation field superposed to a Schwarzschild or Kerr spacetime on test particle motion was investigated. 
Scattering (absorbing and consequent re-emitting) of such radiation by moving particles causes in this case a drag force 
which acts on the particles determining deviations from geodesic motion, termed as Poynting-Robertson effect.
For instance, for a body initially in a circular orbit, there are two kinds of solutions: those in which the body spirals inward or spirals outward, depending on the strength of the radiation pressure.
The former results of Robertson \cite{Robertson-37} are obtained from the Schwarzschild case in the weak field and slow motion approximation for a small drag coefficient. 
The first example in which the Poynting-Robertson effect has been investigated in a self-consistent way, i.e., without the requirement that the radiation field itself be a test field, is that considered in Ref. \cite{vaidyaPR}, where the Vaidya spacetime has been taken as the background spacetime and a Thomson-like interaction of its null dust with test particle motion was assumed. 

Let the force acting on a massive particle with $4$-velocity $U$ be proportional to the momentum density of the radiation field as seen in the test particle rest space, i.e.,
\beq
\label{frad}
f_{\rm (rad)}{}_\alpha=-\sigma P(U)_{\alpha \beta} T^\beta{}_\mu U^\mu\,,
\eeq 
where $P(U)=g+U\otimes U$ projects orthogonally to $U$, $\sigma$ is the effective interaction cross section modeling the absorption and consequent re-emission of radiation by the particle and $T$ is the energy-momentum tensor (\ref{Tmunu}) source of the spacetime.
Note that $\sigma$ has the dimension of a length squared and is assumed to be a constant. 
The equations of motion thus write as 
\beq
M a(U)=f_{\rm (rad)}\,,
\eeq
where $a(U)=\rmD U/\rmd\tau$ denotes the $4$-acceleration of $U$, $U$ is given by Eq. (\ref{Udef}) and $M$ is the particle's rest mass.
An explicit calculation shows
\begin{eqnarray}
f_{\rm (rad)}&=&2 \sigma \omega^2 \gamma (1-\nu^{\hat z})\left\{ [\gamma^2 (1-\nu^{\hat z})-1]m+\gamma^2 (1-\nu^{\hat z})(\nu^{\hat x}e_{\hat x}+\nu^{\hat y}e_{\hat y})
-[\gamma^2\nu^{\hat z}(\nu^{\hat z}-1)+1]e_{\hat z} \right\}\,,
\end{eqnarray}
so that the equations of motion become
\begin{eqnarray}
\label{eqmoto}
\frac{\rmd \nu^{\hat x}}{\rmd \tau} &=& 2\tilde\sigma\omega^2(1-\nu^{\hat z})\nu^{\hat x} -\omega \gamma \nu^{\hat x}\tan s \,(\nu^{\hat x}{}^2+\nu^{\hat y}{}^2+\nu^{\hat z}-1)\,,\nonumber\\
\frac{\rmd \nu^{\hat y}}{\rmd \tau} &=& 2\tilde\sigma\omega^2 (1-\nu^{\hat z})\nu^{\hat y}-\omega\gamma \nu^{\hat y}\tan s \,(\nu^{\hat x}{}^2+\nu^{\hat y}{}^2+\nu^{\hat z}-1)\,,\nonumber\\
\frac{\rmd \nu^{\hat z}}{\rmd \tau} &=& -2\tilde\sigma\omega^2 (1-\nu^{\hat z})^2+\omega\gamma\tan s \,(\nu^{\hat y}{}^2+\nu^{\hat x}{}^2) (1-\nu^{\hat z})\,,
\end{eqnarray}
which must be completed with the evolution equations for $t$, $x$, $y$ and $z$ (see Eq. (\ref{Ucompts})), i.e.,
\beq
\label{eqevol}
\frac{\rmd t}{\rmd \tau}=\gamma\,,\qquad 
\frac{\rmd x}{\rmd \tau}=\frac{\gamma\nu^{\hat x}}{\cos s}\,,\qquad 
\frac{\rmd y}{\rmd \tau}=\frac{\gamma\nu^{\hat y}}{\cos s}\,,\qquad 
\frac{\rmd z}{\rmd \tau}=\gamma \nu^{\hat z}\,,
\eeq
where $\tilde \sigma=\sigma/M$ and the simplified notation $\gamma (U,m)\equiv\gamma$ and $\nu(U,m)^{\hat a}\equiv\nu^{\hat a}$ has been used.

Let us consider first the set of equations (\ref{eqmoto}) for the frame components of the spatial velocity.
It is convenient to express the proper time derivative in terms of the derivative with respect to the variable $s$ introduced above according to
\beq
\rmd s=\omega \rmd (t-z)=\omega \gamma (1-\nu^{\hat z})\rmd \tau\,,
\eeq
so that the system becomes
\begin{eqnarray}
\frac{\rmd \nu^{\hat x}}{\rmd s} &=& \frac{2}{\gamma}\tilde\sigma\omega \nu^{\hat x} - \frac{\nu^{\hat x}}{1-\nu^{\hat z}} (\nu^{\hat x}{}^2+\nu^{\hat y}{}^2+\nu^{\hat z}-1)\tan s\,,\nonumber\\
\frac{\rmd \nu^{\hat y}}{\rmd s} &=& \frac{2}{\gamma}\tilde\sigma\omega \nu^{\hat y}- \frac{ \nu^{\hat y}}{1-\nu^{\hat z}}(\nu^{\hat x}{}^2+\nu^{\hat y}{}^2+\nu^{\hat z}-1)\tan s\,,\nonumber\\
\frac{\rmd \nu^{\hat z}}{\rmd s} &=& -\frac{2}{\gamma}\tilde\sigma\omega  (1-\nu^{\hat z}) +   (\nu^{\hat y}{}^2+\nu^{\hat x}{}^2)  \tan s\,.
\end{eqnarray}
The first two equations imply 
\beq
\nu^{\hat y}\frac{\rmd \nu^{\hat x}}{\rmd s}-\nu^{\hat x}\frac{\rmd \nu^{\hat y}}{\rmd s}=0\,, 
\eeq
whence 
\beq
\frac{\nu^{\hat y}}{\nu^{\hat x}}=const=C\,,
\eeq
so that the set of equations simplifies to
\begin{eqnarray}
\label{fin12}
\frac{\rmd \nu^{\hat x}}{\rmd s} &=& \frac{2}{\gamma}\tilde\sigma\omega \nu^{\hat x} - \frac{\nu^{\hat x}}{1-\nu^{\hat z}} [\nu^{\hat x}{}^2(1+C^2)+\nu^{\hat z}-1]\tan s\,,\nonumber\\
\frac{\rmd \nu^{\hat z}}{\rmd s} &=& -\frac{2}{\gamma}\tilde\sigma\omega  (1-\nu^{\hat z}) +   (1+C^2)\nu^{\hat x}{}^2\tan s\,.
\end{eqnarray}
Eqs. (\ref{fin12}) admit the following first integral
\beq
\frac{\nu^{\hat x}\cos s}{1-\nu^{\hat z}}=const.=B\,.
\eeq
The only surviving equation is thus the following
\beq
\label{eqnuz}
\frac{\rmd \nu^{\hat z}}{\rmd s} = -2\tilde \sigma\omega\frac{(1-\nu^{\hat z})^{3/2}}{\cos s}\left[\nu^{\hat z}(R^2+\cos^2s)-R^2+\cos^2s\right]^{1/2}+\frac{R^2(1-\nu^{\hat z})^2\sin s}{\cos^3s}\,,
\eeq
where $R=B\sqrt{1+C^2}$.
Next setting   
\beq
\nu^{\hat z}=1-\frac{2\cos^2s}{R^2+2W(s)\cos^2s}\,,
\eeq
and substituting then in Eq. (\ref{eqnuz}) leads to the following equation for $W(s)$
\beq
\frac{\rmd W}{\rmd s} =-2\tilde \sigma\omega\sqrt{2W-1}\,,
\eeq
whose general solution is 
\beq
W(s)=\frac{1}{2}+\frac{1}{2}(\sqrt{2W(0)-1}-2\tilde \sigma\omega s)^2\,,
\eeq
which exhibits a parabolic behavior with its minimum $W(0)$ at $s=0$.

Summarizing, the solutions for the frame components of the spatial velocity are 
\begin{eqnarray}
\label{nusolfin}
\nu^{\hat x}&=&\frac{2B\cos s}{R^2+2W(s)\cos^2s}\,, \nonumber\\
\nu^{\hat y}&=&\frac{2BC\cos s}{R^2+2W(s)\cos^2s}\,, \nonumber\\
\nu^{\hat z}&=&1-\frac{2\cos^2s}{R^2+2W(s)\cos^2s}\,,
\end{eqnarray}
where the arbitrary constants $B$, $C$ and $R$ can be in turn expressed in terms of the initial values as 
\beq
B=\frac{\nu^{\hat x}(0)}{1-\nu^{\hat z}(0)}\,,\qquad
C=\frac{\nu^{\hat y}(0)}{\nu^{\hat x}(0)}\,,\qquad R=\frac{\sqrt{\nu^{\hat x}(0)^2+\nu^{\hat y}(0)^2}}{1-\nu^{\hat z}(0)}\,,
\eeq
with 
\beq
W(0)=\frac{1}{1-\nu^{\hat z}(0)}-\frac{R^2}{2}\,.
\eeq
When $\tilde\sigma=0$ (geodesic case) the solution is still given by Eq. (\ref{nusolfin}) with $W(s)=W(0)$ and coincides with Eq. (\ref{geo}).

\begin{figure} 
\typeout{*** EPS figure 1}
\begin{center}
$\begin{array}{cc}
\includegraphics[scale=0.3]{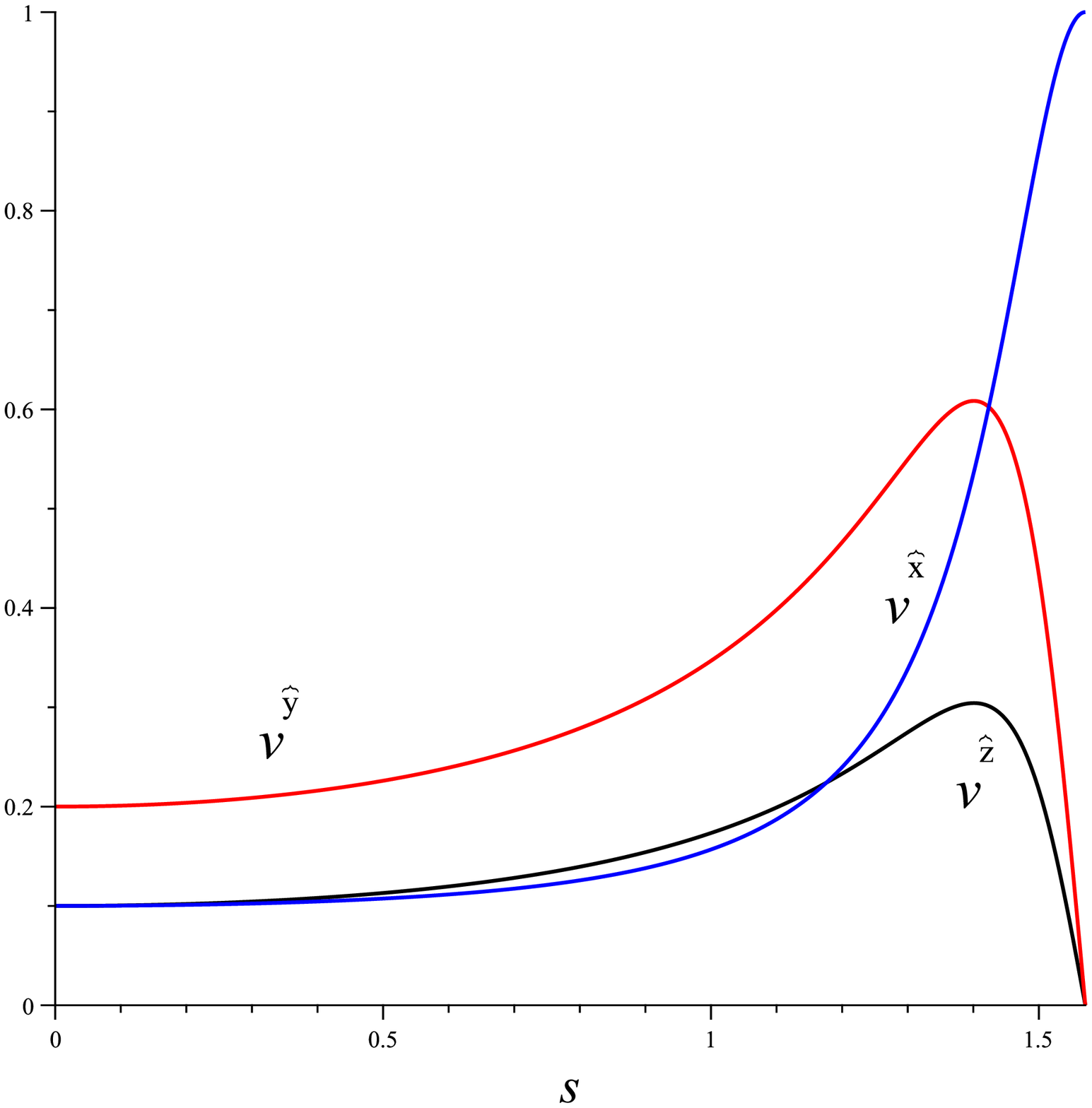}&\quad
\includegraphics[scale=0.3]{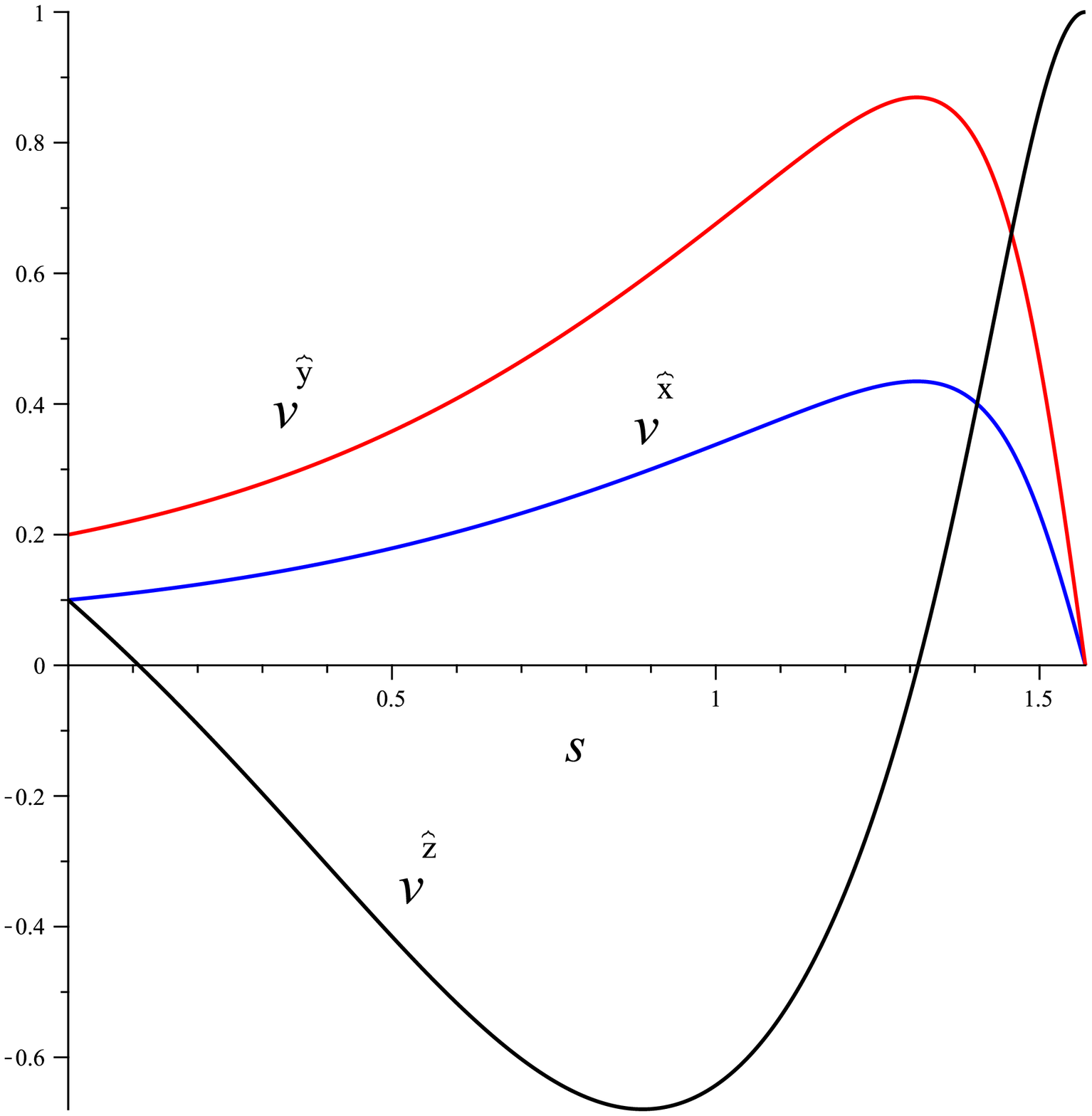}\\[.4cm]
\quad\mbox{(a)}\quad &\quad \mbox{(b)}\\
\end{array}$
\end{center}
\caption{The behavior of the frame components of the spatial velocity is shown as functions of $s=\omega(t-z)\in[0,\pi/2]$ with initial conditions $\nu^{\hat x}(0)=0.1$, $\nu^{\hat y}(0)=0.2$ and $\nu^{\hat z}(0)=0.1$. 
Fig. (a) corresponds to the geodesic case $(\tilde\sigma=0)$, Fig. (b) to the case $\omega\tilde\sigma=0.5$.
Note that in the non-geodesic case the behavior of the $z$-component of the spatial velocity is different with respect to the geodesic case. From Fig. (b), in fact, we see that $\nu^{\hat z}(s)$ decreases and become negative, even for a large interval of the  variable $s$ if  $\tilde\sigma$ is large enough, before turning again positive as approaching its maximum value $1$ for $s=\pi/2$.
Further increase of $\tilde\sigma$ causes practically $\nu^{\hat x}(s)$ and $\nu^{\hat y}(s)$ to vanish and $\nu^{\hat z}(s)$ to approach $1$, after a small interval of values of $s$, in which a transient oscillating behavior is still present.
}
\label{fig:1}
\end{figure}

Fig. \ref{fig:1} shows the behavior of the frame components of the spatial velocity as functions of $s$.
Note the relations 
\beq
\lim_{s\to \pi/2}\nu^{\hat x}=0\,, \quad
\lim_{s\to \pi/2}\nu^{\hat y}=0\,, \quad
\lim_{s\to \pi/2}\nu^{\hat z}=1\,,
\eeq
and
\begin{eqnarray}
&& |\nu^{\hat x}|\le \frac{2B}{R^2}\,,\quad |\nu^{\hat y}|\le \frac{2BC}{R^2}\,,\quad |1-\nu^{\hat z}|\le \frac{2}{R^2}\,,
\end{eqnarray}
which are also evident from the plots. 

In the special case of motion along the $z$ axis, i.e., $\nu^{\hat x}(0)=0=\nu^{\hat y}(0)$, the solution reduces to
\beq
\nu^{\hat x}=0\,, \quad 
\nu^{\hat y}=0\,, \quad
\nu^{\hat z}=1-\frac{1}{W(s)}\,,
\eeq
which also explains the meaning of the function $W$ introduced above. 
If instead the initial conditions are $\nu^{\hat y}(0)=0=\nu^{\hat z}(0)$, $\nu^{\hat x}(0)>0$ (motion starting along the positive $x$-axis), we have
\beq
B=\nu^{\hat x}(0)=R\,,\quad C=0\,,\quad W(0)=1-\frac{\nu^{\hat x}(0)^2}{2}\,,
\eeq
whence
\begin{eqnarray}
\label{nusolfin_x}
\nu^{\hat x}=\frac{2\nu^{\hat x}(0) \cos s}{\nu^{\hat x}(0)^2+2W(s)\cos^2s}\,, \quad
\nu^{\hat y}=0\,, \quad
\nu^{\hat z}=1-\frac{2\cos^2s}{\nu^{\hat x}(0)^2+2W(s)\cos^2s}\,.
\end{eqnarray}

Eqs. (\ref{nusolfin}) can be further integrated to obtain the solution for the accelerated orbit, whose equations (\ref{eqevol}) in terms of the variable $s$ write as
\begin{eqnarray}
\frac{\rmd t}{\rmd s} &=& \frac{1}{\omega}\frac{1}{1-\nu^{\hat z}}=\frac{1}{\omega}\left[\frac{R^2}{2\cos^2s}+W(s)\right]\,,\nonumber\\
\frac{\rmd x}{\rmd s} &=& \frac{1}{\omega\cos s}\frac{\nu^{\hat x}}{1-\nu^{\hat z}}=\frac{1}{\omega}\frac{B}{\cos^2s}\,,\nonumber\\
\frac{\rmd y}{\rmd s} &=& \frac{1}{\omega\cos s}\frac{\nu^{\hat y}}{1-\nu^{\hat z}}=\frac{1}{\omega}\frac{BC}{\cos^2s}\,,\nonumber\\
\frac{\rmd z}{\rmd s} &=& \frac{1}{\omega}\frac{\nu^{\hat z}}{1-\nu^{\hat z}}=\frac{1}{\omega}\left[\frac{R^2}{2\cos^2s}+W(s)-1\right]\,.
\end{eqnarray} 
The integration is straightforward
\begin{eqnarray}
t(s) &=& \frac{W(0)}{\omega}s-\tilde\sigma s^2\left[\sqrt{2W(0)-1}-\frac23\omega\tilde\sigma s\right]+\frac{R^2}{2\omega}\tan s + t_0\,,\nonumber\\
x(s) &=& \frac{B}{\omega}\tan s + x_0\,,\nonumber\\
y(s) &=& \frac{BC}{\omega}\tan s + y_0\,,\nonumber\\
z(s) &=& \frac{W(0)-1}{\omega}s-\tilde\sigma s^2\left[\sqrt{2W(0)-1}-\frac23\omega\tilde\sigma s\right]+\frac{R^2}{2\omega}\tan s + z_0\,,
\end{eqnarray}
with $x_0^\alpha=x^\alpha(0)$.

\begin{figure} 
\typeout{*** EPS figure 2}
\begin{center}
\includegraphics[scale=0.3]{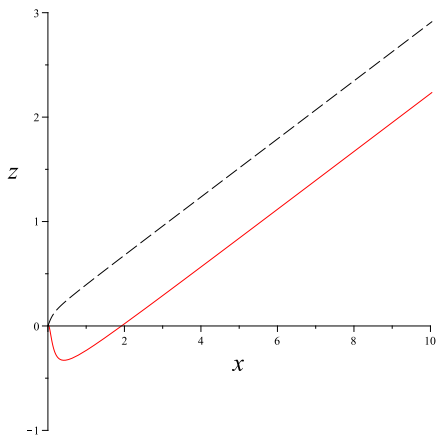}
\end{center}
\caption{The projection of the accelerated orbit on the $x-z$ plane is shown for the same choice of initial conditions for the velocity components as in Fig. \ref{fig:1} and $x(0)=0=z(0)$. 
The dashed curve corresponds to the geodesic case $(\tilde\sigma=0)$, whereas the solid one to the case $\omega\tilde\sigma=0.5$.
}
\label{fig:2}
\end{figure}

Fig. \ref{fig:2} shows the deviation from geodesic motion due to such \lq\lq Poynting-Robertson-like'' effect in a simple 2-dimensional case.


\section{Scattering of massless test fields}

In this section we will use Newman-Penrose formalism to study the scattering of massless test fields by the background radiation field.
We thus switch to the metric signature $+$ $-$ $-$ $-$ to follow standard conventions (we refer to Ref. \cite{chandra} for notations and conventions).
A Newman-Penrose complex null frame can be build up with the orthonormal frame (\ref{ort_frame}), namely
\beq
l=\frac{1}{\sqrt{2}}(m+e_{\hat z})\,,\quad
n=\frac{1}{\sqrt{2}}(m-e_{\hat z})\,,\quad
m=\frac{1}{\sqrt{2}}(e_{\hat x}+i e_{\hat y})\,.
\eeq
It is a principal frame (all the Weyl scalars vanish since the metric is conformally flat) and the only non-vanishing spin coefficient is $\mu=-\sqrt{2}\omega \tan s$.
The background spacetime corresponds to the Faraday tensor 
\beq
F=-{}^{(B)}\phi_2^*[l\wedge m]-{}^{(B)}\phi_2[l\wedge m^*] 
=2\omega\, l\wedge\omega^{\hat x}\,, \qquad
\omega^{\hat x}=\frac{1}{\cos s}\rmd x\,,
\eeq
and associated potential 
\beq
A=\sqrt{2}\ln\left(\frac{1+\sin s}{\cos s}\right)\rmd x\,.
\eeq
The invariants of this field all vanish
\beq
I=16\left[{}^{(B)}\phi_0{}^{(B)}\phi_2-{}^{(B)}\phi_1^2\right]=0\,,
\eeq
implying that it corresponds to a singular (or radiation) field according to the standard classification, as already known.

The equations for perturbing fields of any spin weight $w$ can be summarized by the following set {\it \`a la} Teukolsky \cite{teuk}
\begin{eqnarray}
0&=&[D(\Delta+\mu)-\delta\delta^*]\psi\,, \hspace{7.5cm} (w=1/2, 1)\,,\nonumber\\
0&=&\{[\Delta +\mu^*-2w\mu]D-\delta^*\delta]\}\psi
=\{[\Delta+(1-2w)\mu]D-\delta^*\delta\}\psi\,, \qquad\qquad (w=-1/2, -1)\,,\nonumber\\
0&=&[D\Delta+\Delta D-\delta\delta^*-\delta^*\delta+(\mu+\mu^*)D]\psi
=2[D\Delta-\delta\delta^*+\mu D]\psi\,, \qquad\,\, (w=0)\,,
\end{eqnarray}
where we have used the results $\mu=\mu^*$, $[D,\Delta]=0=[\delta,\delta^*]$, $D\mu=0=\delta\mu$ and $\Delta\mu=-2\omega^2/\cos^2s$.

\subsection{Scattering of a massless scalar field}

Consider a massless scalar field $\psi(t,x,y,z)$ on this background, obeying the Klein-Gordon equation $\Box\psi=0$, i.e.,
\beq
2\omega\tan [\omega(t-z)] (\psi_t+\psi_z)-\psi_{tt}+\frac{1}{\cos^2[\omega(t-z)]}(\psi_{xx}+\psi_{yy})+\psi_{zz}=0\,.
\eeq
The general solution of this equation obtained by separation of variables is
\beq
\label{KGsol}
\psi=\frac{A}{\cos s}e^{i\Phi}\,,
\eeq
where
\beq
\label{phi_def}
\Phi=\Phi_\omega (k_u,k_x,k_y; x^\alpha)\equiv -\frac{k_u}{\sqrt{2}\omega} r-\frac{k_x^2+k_y^2}{2\sqrt{2}\omega k_u}\tan s+k_xx+k_yy\,,
\eeq
and 
\beq
\label{trasfrs}
s=\omega(t-z)\,, \quad
r=\omega(t+z)\,. 
\eeq
The phase factor $\Phi$ has the relevant property that $\partial_\alpha\Phi=K_{\rm (g)}{}_\alpha$ is the null  geodesic defined in Eq. (\ref{null_geo}).
The scaling factor $(\cos s)$ is exactly $[-{\rm det}\, g]^{1/4}$.
Therefore the solution (\ref{KGsol}) can be written as
\beq
\psi=\frac{A}{[-{\rm det}\, g]^{1/4}}e^{i\int K_\alpha\rmd x^\alpha}\,.
\eeq
Clearly, the constant-$\Phi$ surfaces are not hyperplanes. 
In fact, deviations from a planar behavior corresponding to the limiting case $\omega=0$, i.e.,
\beq
\Phi_0=-\frac{k_u}{\sqrt{2}\omega} r-\frac{k_x^2+k_y^2}{2\sqrt{2}\omega k_u}s+k_xx+k_yy\,,
\eeq
are measured by the quantity
\beq
\Phi_\omega-\Phi_0=\frac{k_x^2+k_y^2}{2\sqrt{2}\omega k_u}(s-\tan s)
\simeq  - \frac{k_x^2+k_y^2}{2\sqrt{2}\omega k_u} \frac{s^3}3 \left(1+\frac{2s^2}{5}+O(6)\right) \,,
\eeq
which is nonzero as soon as $s$ increases, whereas the planar regime maintains up to second order in $s$ in the series expansion around $s=0$.
As we explicitly did for null geodesics in Section II, it is also worth to analyze the special case $k_u=\Omega/\sqrt{2}$, $k_y=0$ and $k_x=\Omega$.
We find for the phase factor before and in presence of the wave
\beq
\label{phi_x}\Phi_0 (\Omega/\sqrt{2},\Omega,0; x^\alpha)= - \Omega (t-x) \,,\qquad
\Phi_\omega (\Omega/\sqrt{2},\Omega,0; x^\alpha)= -\frac{\Omega}{2\omega} r-\frac{\Omega}{2\omega }\tan s+\Omega x  
\,,
\eeq
showing the dependence on the variable $z$ as a new feature.

\subsection{Scattering of a massless spin $1/2$ field}

Let us consider the case of a massless Dirac particle, whose spinor components of the corresponding wave function satisfy the following set of first order equations \cite{chandra}
\begin{eqnarray}
&& D F_1+\delta^*F_2=0\,,\quad 
(\Delta +\mu)F_2+\delta F_1=0\,,\nonumber\\
&& D G_2-\delta G_1=0\,,\quad 
(\Delta +\mu^*)G_1-\delta^* G_2=0\,.
\end{eqnarray}
We have then $G_1=-F_2^*$ and $G_2=F_1^*$ so that the only equations to be solved are
\begin{eqnarray}
D F_1+\delta^*F_2=0\,,\quad 
(\Delta +\mu)F_2+\delta F_1=0\,.\nonumber\\
\end{eqnarray}
Looking for solutions of the form
\beq
F_1=A_1(s)e^{i\Phi}\,,\qquad F_2=A_2(s)e^{i\Phi}\,,
\eeq
we find
\beq
A_1(s)=\frac{A_1}{\cos^2 s}\,, \qquad
A_2(s)=\frac{A_2}{\cos s}\,, \qquad
A_1=\frac{k_-}{\sqrt{2}k_u}A_2\,.
\eeq

\subsection{Scattering of a source-free test electromagnetic field}

Let us consider now a superposed electromagnetic field to the background.
Maxwell's equations for the electromagnetic complex quantities $\phi_0$, $\phi_1$ and $\phi_2$ (to be distinguished from the background ones) are listed in Ref. \cite{chandra} and in this case reduce to
\begin{eqnarray}
D\phi_1-\delta^*\phi_0=0\,,\quad 
D\phi_2-\delta^*\phi_1=0\,,\quad
\delta \phi_1-\Delta \phi_0=\mu \phi_0\,,\quad
\delta \phi_2-\Delta \phi_1=2\mu \phi_1\,.
\end{eqnarray}
The solution of these equations is straightforward, namely
\begin{eqnarray}
\phi_0=\frac{A_0}{\cos s}e^{i\Phi}\,,\qquad \phi_1=\frac{A_1}{\cos^2 s}e^{i\Phi}\,,\qquad
\phi_2=\frac{A_2}{\cos^3 s}e^{i\Phi}\,,
\end{eqnarray}
where $\Phi$ is defined in Eq. (\ref{phi_def}) and with
\beq
A_0=-\frac{\sqrt{2}k_u}{k_-}A_1\,,\qquad A_2=-\frac{k_-}{\sqrt{2}k_u}A_1\,,\qquad k_\pm =k_x\pm ik_y\,,
\eeq
and $A_1$ and arbitrary (complex) constant.
The invariants of this field vanish 
\beq
I=16\left(\phi_0\phi_2-\phi_1^2\right)=0\,,
\eeq
so it corresponds again to a singular (or radiation) field.

Taking into account that $\phi_0$ has spin weight $w =1$, $\phi_1$ has spin weight $w=0$ and $\phi_2$ has spin weight $w=-1$ one may summarize the solution in a single formula
\beq
\phi_w=\frac{A_w}{\cos^{2-w} s}e^{i\Phi}\,.
\eeq

\subsubsection{The dielectric medium analogy}

It is well known \cite{skro,pleb,volkov} that the equations of electromagnetic waves in a gravitational field can be interpreted as the Maxwell's equations in flat spacetime but in a material medium characterized by electric (equal to magnetic) permittivity defined in terms of the metric coefficients.
A lot of applications of this formalism have been discussed in different backgrounds by B. Mashhoon (see, e.g., Refs. \cite{mas73,mas75}).
This approach (perfectly equivalent either to the Newman-Penrose approach developed above or to any other observer and frame dependent analysis) allows to deal with \lq\lq flat spacetime quantities," a fact that facilitates their interpretation.  
In this framework a Cartesian coordinate system is introduced so that the electric and magnetic fields are defined using the decomposition $F_{\mu\nu}\to ({\mathbf E},{\mathbf B})$ and $\sqrt{-g}F^{\mu\nu}\to (-{\mathbf D},{\mathbf H})$.
Maxwell's equations turn out to be formally equivalent to the electromagnetic field equations in a dielectric medium in flat spacetime together with the constitutive relations
\beq
D_i=\epsilon_{ij}E_j -(G\times H)_i\,,\qquad B_i=\mu_{ij}H_j+(G\times E)_i\,,
\eeq
where
\beq
\epsilon_{ij}=\mu_{ij}=-\sqrt{-g}\, \frac{g^{ij}}{g_{tt}}\,,\qquad G_i=-\frac{g_{0i}}{g_{tt}}\,.
\eeq
In the present case $g_{0i}=0$, $g_{tt}=1$, $\sqrt{-g}=\cos^2 s$, so that the above relations become
\beq
D_i=\epsilon_{ij}E_j\,,\qquad B_i=\mu_{ij}H_j\,,
\eeq
with 
\beq
\epsilon_{ij}=\mu_{ij}=\cos^2 s\, g^{ij}={\rm diag}[1,1,\cos^2 s]\,,\qquad G_i=0\,,
\eeq
representing a medium which in a sense is optically active only along the $z$ direction.
One then introduces the two complex vectors
\beq
{\mathbf F}^\pm ={\mathbf E}\pm i {\mathbf H}\,,\qquad {\mathbf S}^\pm ={\mathbf D}\pm i {\mathbf B}\,;
\eeq
the constitutive relations in this case are summarized by $S^\pm_i=\epsilon_{ij}F^\pm_j$.

It is possible and convenient to write the electromagnetic equations in a form which is particularly suitable for the discussion of wave phenomena, namely
\beq
\frac{1}{i}\nabla \times {\mathbf F}^\pm =\pm \partial_t {\mathbf S}^\pm\,,\qquad \nabla \cdot {\mathbf S}^\pm=0\,.
\eeq
Looking for solutions of the form
\beq
{\mathbf F}^\pm =C_\pm\bfphi^\pm(s) e^{i\Phi_\pm}\,,
\eeq
with $C_\pm$ arbitrary constants, gives
\beq
\label{Phi_pm}
\Phi_\pm=-\frac{k_u^\pm}{\sqrt{2}\omega}r-\frac{(k_x^\pm)^2+(k_y^\pm)^2}{2\sqrt{2}\omega k_u^\pm}\tan s+k_x^\pm x+k_y^\pm y
=\Phi_\omega(k_u^\pm,k_x^\pm, k_y^\pm; x^\alpha)\,,
\eeq
where the variables $r$ and $s$ are defined by Eq. (\ref{trasfrs}) (see also Eq. (\ref{phi_def})), whereas
\beq
\bfphi^+ \equiv 
\left(
\begin{array}{c}
\phi_1^+ \\
\phi_2^+ \\
\phi_3^+ 
\end{array}
\right)
=
\frac{1}{\cos^2 s}
\left(
\begin{array}{c}
1 \\
i \\
-\frac{2\sqrt{2}k_u^+}{k_x^+-ik_y^+}
\end{array}
\right)
-\frac{2k_u^+{}^2}{(k_x^+-ik_y^+)^2}
\left(
\begin{array}{c}
1 \\
-i \\
0 
\end{array}
\right)
\eeq
and $\bfphi^-=(\bfphi^+)^*$ with $(k_u^+,k_x^+,k_y^+)\to(k_u^-,k_x^-,k_y^-)$.
For instance, the components of $\mathbf F$ in the $+$ polarized case are
\begin{eqnarray}
F^+_x &=&E_x+iH_x=C_+ \phi^+_1(s) e^{i\Phi_+}= C_+\left[\frac{1}{\cos^2 s}- \frac{2(k_u^+)^2(k_x^+{}^2-k_y^+{}^2+2ik_x^+k_y^+)}{(k_x^+{}^2+k_y^+{}^2)^2}\right](\cos \Phi_++i \sin \Phi_+)\,, \nonumber\\
F^+_y &=&E_y+iH_y=C_+ \phi^+_2(s) e^{i\Phi_+}= iC_+\left[\frac{1}{\cos^2 s}+ \frac{2(k_u^+)^2(k_x^+{}^2-k_y^+{}^2+2ik_x^+k_y^+)}{(k_x^+{}^2+k_y^+{}^2)^2}\right](\cos \Phi_++i \sin \Phi_+)\,, \nonumber\\
F^+_z &=&E_z+iH_z=C_+ \phi^+_3(s) e^{i\Phi_+}= -\frac{C_+}{\cos^2 s} \frac{2\sqrt{2}k_u^+}{(k_x^+{}^2+k_y^+{}^2)}(k_x^++ik_y^+) [\cos \Phi_++i \sin \Phi_+]\,.
\end{eqnarray}
The above relations can be simplified if one consider the constants $k_u^+ $, $k_x^+ $ and $k_y^+$  (and similarly in case $-$) related by
\beq
k_u^+=\frac{\Omega}{\sqrt{2}}\,,\qquad k_x^+=\Omega \cos \alpha \,,\qquad k_y^+=\Omega \sin \alpha\,,
\eeq
corresponding in the flat spacetime limit to a circular polarization for a wave propagating along the direction $x\cos \alpha + y\sin \alpha $ on the $x-y$ plane.
With the above choice of constants we have
\beq
\Phi_+=-\Omega \left[\frac12 (t+z)+\frac{\tan \omega (t-z)}{2\omega} -\cos\alpha x -\sin \alpha y \right]\,,\qquad
\bfphi^+
=
\frac{1}{\cos^2 s}
\left(
\begin{array}{c}
1 \\
i \\
-2e^{i\alpha}
\end{array}
\right)
-e^{2i\alpha}
\left(
\begin{array}{c}
1 \\
-i \\
0 
\end{array}
\right)\,,
\eeq
so that
\beq
\bfphi^+ e^{i\Phi_+}=\frac{1}{\cos^2 s}
\left(
\begin{array}{c}
e^{i\Phi_+} \\
e^{i(\Phi_++\pi/2)} \\
-2e^{i(\alpha+\Phi_+)}
\end{array}
\right)
-
\left(
\begin{array}{c}
e^{i(2\alpha+\Phi_+)} \\
-e^{i(2\alpha+\Phi_++\pi/2)} \\
0 
\end{array}
\right)\,.
\eeq

Let us consider as an example the special case $\alpha=0$, corresponding in the flat spacetime limit to a circularly polarized wave propagating along the $x$ direction.
The complex vector ${\mathbf F}^+$ thus simplifies to
\beq
{\mathbf F}^+=
C_+
\left(
\begin{array}{c}
\left(\displaystyle \frac{1}{\cos^2 s}-1\right)e^{i\Phi_+} \\
\left(\displaystyle\frac{1}{\cos^2 s}+1\right)e^{i(\Phi_++\pi/2)} \\
-\displaystyle\frac{2}{\cos^2 s} e^{i\Phi_+}
\end{array}
\right)\,, \qquad
\Phi_+=\Omega\left[-\frac12(t+z)-\frac{1}{2\omega}\tan[\omega(t-z)]+ x\right]\,,
\eeq
with associated electric and magnetic field components given by
\begin{eqnarray}
\label{EHxyz_ex}
E_x(x^\alpha)&=& C_+\left(\frac{1}{\cos^2 s}-1\right)\cos \Phi_+\,,\quad
E_y(x^\alpha)= -C_+\left(\frac{1}{\cos^2 s}+1\right)\sin \Phi_+\,,\quad
E_z(x^\alpha)= -C_+\frac{2}{\cos^2 s}\cos \Phi_+\,,\nonumber\\
H_x(x^\alpha)&=& C_+\left(\frac{1}{\cos^2 s}-1\right)\sin \Phi_+\,, \quad
H_y(x^\alpha)= C_+\left(\frac{1}{\cos^2 s}+1\right)\cos \Phi_+\,, \quad
H_z(x^\alpha)= -C_+\frac{2}{\cos^2 s}\sin \Phi_+\,.
\end{eqnarray}
Since we are interested in determining the modification of the incoming electromagnetic field when the interaction with the backgound electromagnetic wave is turned on it is convenient to expand the field components in powers of $\omega$. 
A series expansion of Eq. (\ref{EHxyz_ex}) in $\omega$ (truncated to the second order), taking into account that 
\beq
\Phi_+=-\Omega(t-x)-\frac16\omega^2\Omega(t-z)^3+O(\omega^3)\,,
\eeq
thus gives
\beq
\label{EHexp}
{\mathbf E}={\mathbf E}^0+\omega^2 (t-z)^2{\mathcal D}_{\rm{(E)}}{\mathbf E}^0+O(\omega^3)\,, \qquad
{\mathbf H}={\mathbf H}^0+\omega^2 (t-z)^2{\mathcal D}_{\rm{(H)}}{\mathbf E}^0+O(\omega^3)\,, 
\eeq
where 
\begin{eqnarray}
{\mathbf E}^0&=&(0,E_y^0,E_z^0)\,, \qquad
E_y^0=2C_+\sin[\Omega(t-x)]\,,\qquad
E_z^0=-2C_+\cos[\Omega(t-x)]\,, \nonumber\\
{\mathbf H}^0&=&\tilde C\,{\mathbf E}^0\,, \qquad
\tilde C=
\left( \begin{array}{ccc}
0 & 0 & 0 \\
0 & 0 & -1 \\
0 & 1 & 0 \\
\end{array}
\right)\,,
\end{eqnarray}
and the ``scattering'' matrices ${\mathcal D}_{\rm{(E)}}$ and ${\mathcal D}_{\rm{(H)}}$ are given by
\beq
{\mathcal D}_{\rm{(E)}}=-{\mathcal D}_{\rm{(H)}}\tilde C=
\left( \begin{array}{ccc}
0 & 0 & -\frac12 \\
0 & \frac12 & -\frac16 \Omega (t-z) \\
0 & \frac16 \Omega (t-z) & 1 \\
\end{array}
\right)\,. 
\eeq
As a result, the interaction with the background electromagnetic wave has changed the incoming monocromatic electromagnetic wave with a given frequency and direction of propagation to a more complex field. 

In order to analyze the properties of the scattering matrices, let us rewrite the matrix ${\mathcal D}_{\rm{(E)}}$ in terms of its symmetric and antisymmetric parts as follows
\beq
{\mathcal D}_{\rm{(E)}}=
\left( \begin{array}{ccc}
0 & 0 & -\frac14 \\
0 & \frac12 & 0 \\
-\frac14 &  0 & 1 \\
\end{array}
\right)+
\left( \begin{array}{ccc}
0 & 0 & -\frac14 \\
0 & 0& -\frac16 \Omega (t-z) \\
\frac14 &  \frac16 \Omega (t-z) & 0 \\
\end{array}
\right)\equiv S+A\,.
\eeq
The effects of the interaction can thus be summarized by a rotation (due to the antisymmetric matrix $A$) plus a deformation (due to the symmetric matrix $S$) of the initial electric field. {\it Only} the rotation matrix depends on the frequency $\Omega$ of the incoming field and can be associated with a rotation vector in a standard way, i.e.,
$A^i=\frac12 \epsilon^{ijk}A_{jk}$ ($\epsilon^{ijk}$ being the Levi-Civita indicator of the three dimensional Euclidean space). 
Therefore, the electric field (\ref{EHexp}) can be rewritten in the form
\beq
{\mathbf E}\simeq{\mathbf E}^0+\omega^2 (t-z)^2 \left(S {\mathbf E}^0 + {\mathbf A} \times  {\mathbf E}^0\right)\,,
\eeq
where
\beq
{\mathbf A}=\frac16 \Omega (t-z)e_x -\frac14 e_{y}\,,\qquad 
S {\mathbf E}^0=
\left( 
\begin{array}{c}
-\frac14 E_z^0 \\
\frac12 E_y^0  \\
E_z^0\\
\end{array}
\right)\,,
\eeq
which shows the coupling between the two frequencies $\Omega$ and $\omega$ only in the rotational part of the total effect of the interaction and can be in principle measurable.
Finally, introducing magnitude and direction for ${\mathbf A}$, i.e., ${\mathbf A}=|{\mathbf A}|{\hat \mathbf {a}}$, with
\beq
|{\mathbf A}|=\frac14 \sqrt{\frac{4 \Omega^2}{9}  (t-z)^2+ 1}\,,
\eeq
we have
\beq
{\mathbf E}\simeq\left[{\mathbf I}+(\Delta \beta)\, \hat \mathbf {a}\, \times\,\right]  {\mathbf E}^0 + \omega^2 (t-z)^2  S {\mathbf E}^0 \,,\quad \Delta \beta=\omega^2 (t-z)^2 |{\mathbf A}|\,,
\eeq
implying that ${\mathbf E}$ is obtained from ${\mathbf E}^0$ by a rotation of an angle $\Delta \beta$ around the axis $\hat \mathbf {a}$, plus a deformation effect.

\section{Concluding remarks}

We have studied the interaction of test particles and fields of different spin with the radiation field of an electromagnetic plane wave.
First of all we have integrated the geodesic equations to investigate the motion of test particles.
Then we have considered a more sophisticated interaction with the radiation field leading to accelerated orbits.
The force term entering the equations of motion is taken to be proportional to the 4-momentum density of radiation observed in the particle's rest frame by the effective interaction cross section (assumed to be a constant) modeling the absorption and consequent re-emission of radiation by the particle.
Such an approach has been widely used in the recent literature to study the problem of scattering of test particles undergoing the action of a Thomson-type interaction with a radiation field in different backgrounds of astrophysical interest leading to the well known Poynting-Robertson effect.
We have then analyzed the deviations from geodesic motion due to such \lq\lq Poynting-Robertson-like'' effect, taking advantage from the explicit analytic solutions we have been able to produce.
Finally we have studied the scattering of fields by the electromagnetic wave, i.e., scalar (spin 0), massless spin $\frac12$ and electromagnetic (spin 1), elucidating the different kind of interactions.
The scattering of electromagnetic radiation has been also analysed in terms of the \lq\lq dielectric medium analogy'' of a gravitational field. 
According to this framework, the electromagnetic field equations can be cast into the form of Maxwell's equations in flat spacetime but in a \lq\lq medium'' with prescribed dielectric and permeability tensors.
This method allows to treat electromagnetic phenomena in curved spacetime by using techniques which are familiar from the flat spacetime theory. 
As an example we have considered the case of a superposed monocromatic electromagnetic wave to the background field corresponding in the flat spacetime limit to a circularly polarized wave propagating along the $x$ direction.
As a result of the interaction the incoming wave is changed to a more complex field propagating along a different direction, the electric field undergoing a rotation plus a deformation effect.
We have found that the coupling between the two frequencies $\Omega$ and $\omega$ of the superposed and background electromagnetic wave is responsible for the rotational effect of the interaction and can be in principle measurable.

\begin{acknowledgments}
The authors acknowledge ICRANet for support. 
\end{acknowledgments}


\begin{thebibliography}{00}


\bibitem{Poynting-03}
J. H. Poynting, 
Phil.\ Trans.\ Roy.\ Soc. {\bf 202}, 525 (1904).

\bibitem{Robertson-37}
H. P. Robertson,  
Mon.\ Not.\ R.\ Astron.\ Soc. {\bf 97}, 423 (1937).

\bibitem{BiniJS-09}
D. Bini, R. T. Jantzen, and L. Stella,
Class.\ Quantum Grav. {\bf 26}, 055009 (2009).

\bibitem{BiniGJSS-11}
D. Bini, A. Geralico, R. T. Jantzen, O. Semer\'ak, and L. Stella, 
Class.\ Quantum Grav. {\bf 28}, 035008 (2011).

\bibitem{vaidyaPR}
D. Bini, A. Geralico, R. T. Jantzen, and O. Semerak,
Class.\ Quantum Grav. {\bf 28}, 245019 (2011).

\bibitem{Vaidya-43}
P. C. Vaidya,
Current Sci. (India) {\bf 12}, 183 (1943).

\bibitem{Vaidya-51}
P. C. Vaidya,  
Proc. Indian Acad. Sci. A {\bf 33}, 264 (1951).

\bibitem{grif}
J. B. Griffiths, 
Phys. Lett. A {\bf 54}, 269 (1975).

\bibitem{grif2}
J. B. Griffiths, 
Ann. Phys. (N.Y.) {\bf 102}, 388 (1976).

\bibitem{grif3}
J. B. Griffiths, 
{\it Colliding Plane Waves in General Relativity} (Oxford University Press, Oxford, 1991).

\bibitem{gurtug1}
O. Gurtug, M. Halilsoy, and O. Unver, 
Phys. Rev. D {\bf 74}, 044020 (2006).

\bibitem{gurtug2}
M. Halilsoy and O. Gurtug,
Phys. Rev. D {\bf 75}, 124021 (2007).

\bibitem{book}
F. de Felice and D. Bini, 
\textit{Classical Measurements in Curved Space-times} (Cambridge University Press, Cambridge, 2010).
 
\bibitem{chandra}  
S. Chandrasekhar,
\textit{The Mathematical Theory of Black Holes} (Oxford University Press, Oxford, 1983).

\bibitem{teuk}
S. A. Teukolsky, 
Astrophys. J. {\bf 185}, 635 (1973).

\bibitem{skro}
G. V. Skrotskii,
Dokl. Akad. Nauk SSSR {\bf 114}, 73 (1957) 
[Sov. Phys. - Dokl. {\bf 2}, 226 (1957)].

\bibitem{pleb}
J. Plebanski, 
Phys. Rev. {\bf 118}, 1396 (1960).

\bibitem{volkov}
A. M. Volkov, A. A. Izmest'ev, and G.V. Skrotskii,
Zh. Eksp. Theor. Fiz. {\bf 59}, 1254 (1970) 
[Sov. Phys. - JETP {\bf 32}, 686 (1971)].

\bibitem{mas73}
B. Mashhoon, 
Phys. Rev. D {\bf 8}, 4297 (1973).

\bibitem{mas75}
B. Mashhoon,
Phys. Rev. D {\bf 11}, 2679 (1975).


\end{thebibliography}
\end{document}